\documentclass[iop]{emulateapj}
\usepackage{apjfonts}
\usepackage{natbib}

\newcommand{\chandra}{{\it Chandra\/}}
\newcommand{\xmm}{\hbox{\it XMM-Newton\/}}
\newcommand{\nustar}{{\it NuSTAR\/}}
\newcommand{\flux}{{erg~cm$^{-2}$~s$^{-1}$}}
\newcommand{\lum}{{erg~s$^{-1}$}}

\usepackage{bm}

\begin{document}

\title{Weak Hard \hbox{X-ray} Emission from Broad Absorption Line
Quasars: Evidence for Intrinsic
X-ray Weakness}
\author{
B.~Luo,\altaffilmark{1,2}
W.~N.~Brandt,\altaffilmark{1,2}
D.~M.~Alexander,\altaffilmark{3}
D.~Stern,\altaffilmark{4}
S.~H.~Teng,\altaffilmark{5}
P.~Ar{\'e}valo,\altaffilmark{6,7}
F.~E.~Bauer,\altaffilmark{6,8,9}
S.~E.~Boggs,\altaffilmark{10}
F.~E.~Christensen,\altaffilmark{11}
A.~Comastri,\altaffilmark{12}
W.~W.~Craig,\altaffilmark{13,10}
D.~Farrah,\altaffilmark{14}
P.~Gandhi,\altaffilmark{3}
C.~J.~Hailey,\altaffilmark{15}
F.~A.~Harrison,\altaffilmark{16}
M.~Koss,\altaffilmark{17}
P.~Ogle,\altaffilmark{18}
S.~Puccetti,\altaffilmark{19,20}
C.~Saez,\altaffilmark{21}
A.~E.~Scott,\altaffilmark{1,2}
D.~J.~Walton,\altaffilmark{16}
\& W.~W.~Zhang\altaffilmark{22}}

\altaffiltext{1}{Department of Astronomy \& Astrophysics, 525 Davey Lab,
The Pennsylvania State University, University Park, PA 16802, USA}
\altaffiltext{2}{Institute for Gravitation and the Cosmos, The Pennsylvania State University, University Park, PA 16802, USA}
\altaffiltext{3}{Department of Physics, Durham University, South Road,
Durham DH1 3LE, UK}
\altaffiltext{4}{Jet Propulsion Laboratory, California Institute of Technology,
Pasadena, CA 91109, USA}
\altaffiltext{5}{Observational Cosmology Laboratory, NASA Goddard Space Flight Center, Greenbelt, MD 20771, USA}
\altaffiltext{6}{Instituto de Astrof\'{\i}sica, Facultad de F\'{i}sica, Pontificia Universidad Cat\'{o}lica de Chile, 306, Santiago 22, Chile}
\altaffiltext{7}{Instituto de F\'isica y Astronom\'ia, Facultad de Ciencias, Universidad de Valpara\'iso, Gran Bretana 1111, Playa Ancha, Valpara\'iso, Chile}
\altaffiltext{8}{Millennium Institute of Astrophysics, Vicu\~{n}a Mackenna 4860, 7820436 Macul, Santiago, Chile} 
\altaffiltext{9}{Space Science Institute, 4750 Walnut Street, Suite
205, Boulder, CO 80301, USA}
\altaffiltext{10}{Space Sciences Laboratory, University of California, Berkeley,
CA 94720, USA}
\altaffiltext{11}{DTU Space - National Space Institute, Technical University of
Denmark, Elektrovej 327, 2800 Lyngby, Denmark}
\altaffiltext{12}{INAF---Osservatorio Astronomico di Bologna, Via Ranzani 1,
Bologna, Italy}
\altaffiltext{13}{Lawrence Livermore National Laboratory, Livermore, CA
94550, USA}
\altaffiltext{14}{Department of Physics, Virginia Tech, Blacksburg, VA 24061, USA}
\altaffiltext{15}{Columbia Astrophysics Laboratory, Columbia University, New
York, NY 10027, USA}
\altaffiltext{16}{Cahill Center for Astronomy and Astrophysics,
California Institute
of Technology, Pasadena, CA 91125, USA}
\altaffiltext{17}{Institute for Astronomy, Department of Physics, ETH Zurich, Wolfgang-Pauli-Strasse 27, CH-8093 Zurich, Switzerland}
\altaffiltext{18}{IPAC, California Institute of Technology, Mail Code 220-6, Pasadena, CA 91125, USA}
\altaffiltext{19}{ASDC---ASI, Via del Politecnico, 00133 Roma, Italy}
\altaffiltext{20}{INAF---Osservatorio Astronomico di Roma, via Frascati 33, 00040 Monte Porzio Catone (RM), Italy}
\altaffiltext{21}{Department of Astronomy, University of Maryland, 
College Park, MD 20742, USA}
\altaffiltext{22}{NASA Goddard Space Flight Center, Greenbelt, MD 20771, USA}

\begin{abstract}
We report
\nustar\ observations of a sample of six 
\hbox{X-ray} weak broad absorption line 
(BAL) quasars. These targets, at $z=0.148$--1.223,
are among the
optically brightest and most luminous BAL quasars known at $z<1.3$.
However, their rest-frame $\approx2$~keV luminosities
are 14 to $>330$ times weaker than expected for typical quasars.
Our results from a pilot \nustar\ study of 
two low-redshift BAL quasars, 
a \chandra\ stacking analysis of 
a sample of high-redshift BAL quasars, and 
a \nustar\ spectral analysis of the local BAL quasar Mrk~231
have already suggested
the existence of intrinsically
X-ray weak BAL quasars, i.e., quasars 
not emitting \hbox{X-rays} at the level
expected from their optical/UV emission. 
The aim of the current program 
is to extend the search 
for such extraordinary objects.
Three of the six new targets
are weakly detected by \nustar\ with $\la45$ counts in the \hbox{3--24~keV} 
band, 
and the other three are not detected. 
The hard X-ray (8--24~keV) weakness observed by \nustar\
requires Compton-thick 
absorption if these objects have nominal underlying X-ray emission.
However, a soft stacked effective photon index ($\Gamma_{\rm eff}\approx1.8$) 
for this sample disfavors Compton-thick absorption in general.
The uniform hard X-ray weakness observed by \nustar\ for this 
and the pilot samples selected with $<10$~keV weakness also suggests
that the X-ray weakness is intrinsic in at least some of the targets.
We conclude that the \nustar\ observations have likely
discovered a significant population ($\ga33\%$) of intrinsically
X-ray weak objects among the BAL quasars with significantly weak 
$<10$~keV emission. We suggest that 
intrinsically \hbox{X-ray} weak quasars might be
preferentially observed as BAL quasars.
\end{abstract}

\section{INTRODUCTION}

X-ray emission is considered to be ubiquitous from 
active galactic nuclei (AGNs),
and it is believed to originate from a hot ``corona'' surrounding the inner
accretion disk via Comptonization of disk optical/UV/EUV photons 
\citep[e.g.,][]{Haardt1991}. 
X-ray emission may be enhanced in radio-loud AGNs due to the 
contribution from jets, and
the X-ray weakness observed in some AGNs is generally attributed to absorption.
After excluding radio-loud AGNs and potentially X-ray absorbed AGNs,
a highly significant correlation between the AGN UV luminosity 
(2500~\AA\ monochromatic luminosity, $L_{\rm 2500~{\textup{\AA}}}$)
and the 
\hbox{X-ray-to-optical} power-law slope parameter ($\alpha_{\rm OX}$)\footnote{$\alpha_{\rm OX}$
is defined as
$\alpha_{\rm OX}=-0.3838\log(f_{2500~{\textup{\AA}}}/f_{2~{\rm~keV}})$,
where $f_{2500~{\textup{\AA}}}$ and $f_{2~{\rm~keV}}$
are the \hbox{rest-frame} 2500~\AA\ and 2~keV flux densities.}
has been established across $\approx5$ orders of magnitude in 
UV luminosity \citep[e.g.,][]{Steffen2006,Just2007,Lusso2010}.
This relation highlights apparently 
uniform physical mechanisms at work at the heart of the AGN engine.

One might naturally wonder whether there are
AGNs that are intrinsically X-ray weak, producing much less
X-ray emission than expected from
the \hbox{$\alpha_{\rm OX}$--$L_{\rm 2500~{\textup{\AA}}}$}
relation. One such example is PHL~1811, a very bright quasar
at $z=0.19$ with a $B$-band magnitude of 13.9 
that has been studied extensively 
\citep[e.g.,][]{Leighly2007b,Leighly2007}. It is believed to be
intrinsically X-ray weak by
a factor of \hbox{$\approx30$--100}.
A small sample of Sloan Digital Sky Survey
(SDSS; \citealt{York2000}) 
quasars with similar emission-line properties,
termed ``PHL~1811 analogs'', has also been observed to be
X-ray weak \citep{Wu2011}, although the nature of their
\hbox{X-ray} weakness (intrinsic X-ray weakness or absorption) 
is uncertain. The fraction of PHL 1811 analogs
in the radio-quiet quasar population is small, $\la1.2\%$.
As a first attempt to constrain the fraction of intrinsically X-ray weak
AGNs systematically, \citet{Gibson2008} searched for such objects
among optically selected SDSS
quasars, again excluding radio-loud AGNs and 
potentially X-ray absorbed systems. Their conclusion was that 
such AGNs are rare; e.g., the fraction is $\la2\%$ for AGNs that 
are intrinsically X-ray weak by a factor of 10 or more.
Discovery of intrinsically X-ray weak
AGNs challenges the idea of a universal X-ray
emission mechanism, and studies of such objects should 
provide insights into the nature of the corona.

There is one significant population of AGNs that belongs to the 
category of potentially X-ray absorbed AGNs which has been
excluded in previous
searches for intrinsically X-ray weak AGNs, and this is
broad absorption line (BAL) quasars. 
BAL quasars comprise \hbox{$\approx15\%$} of optically selected quasars
\citep[e.g.,][]{Hewett2003,Trump2006,Gibson2009,Allen2011},
and observationally they are in general \hbox{X-ray}
weak, often due to absorption 
\citep[e.g.,][]{Gallagher2002a,Gallagher2006,Fan2009,Gibson2009}.
In the accretion-disk wind model (see Figure~1 of \citealt{Luo2013},
hereafter L13),
where a radiatively driven wind is launched from the accretion disk at 
$\approx10^{16}$--10$^{17}$~cm \citep[e.g.,][]{Murray1995,Proga2000},
BALs are observed when
the inclination angle is large and the line of sight passes through
the outflowing wind.
In this model, suppression of the nuclear X-ray emission is 
required to prevent the wind from being overionized. 
Some ``shielding'' material, e.g., the shielding gas
as shown in Figure~1 of L13, is usually invoked to provide 
X-ray absorption in BAL quasars, 
which is consistent with the absorption typically observed. 
However, if a BAL quasar
were intrinsically \hbox{X-ray} weak, the wind could also be launched 
successfully with little or no shielding. 

Indeed, there are some BAL quasars with significant
\hbox{X-ray} weakness that cannot be accounted for 
by the apparent \hbox{X-ray} absorption determined using $<10$~keV 
\chandra\ or \xmm\ data (e.g., \citealt{Sabra2001}; L13). 
These are candidates for intrinsically X-ray weak 
AGNs. It is also possible that they are intrinsically X-ray normal but are 
heavily obscured 
($N_{\rm H}\ga5\times10^{23}$~cm$^{-2}$) or even
Compton-thick ($N_{\rm H}\ge1.5\times10^{24}$~cm$^{-2}$), 
so that the observed $<10$~keV spectra are dominated by a
Compton-reflected component. To distinguish between the intrinsic 
X-ray weakness and heavy absorption scenarios, observations of highly
penetrating X-rays in
the $>10$~keV rest-frame band are required. 

In L13, we presented
\nustar\ \citep{Harrison2013} 
\hbox{3--79 keV} observations of a pilot sample of two 
such BAL quasars,
PG~1004+130 ($z=0.241$) and PG~1700+518 (\hbox{$z=0.292$}).
Both objects are surprisingly \hbox{X-ray} weak
in the \nustar\ band, suggesting either intrinsic
X-ray weakness or highly Compton-thick absorption
($N_{\rm H}\approx7\times10^{24}$~cm$^{-2}$). Additionally, a \chandra\ 
stacking analysis in L13 with
the Large Bright Quasar Survey (LBQS) BAL-quasar sample
at high redshift (where \chandra\ probes the rest-frame
\hbox{$\approx1.5$--24~keV} band) revealed an effective 
power-law photon
index of $\Gamma_{\rm eff}=1.6_{-0.5}^{+0.6}$.
This effective photon index 
is fairly soft/steep for a spectrum expected to be absorbed, and it
argues against
Compton-thick absorption in general, which would usually result in
a hard/flat ($\Gamma_{\rm eff}\approx0.5$ with a range of $\approx0\textrm{--}1$) 
spectrum \citep[e.g.,][]{George1991,Comastri2011,Gandhi2014,Rovilos2014}.
This result suggests a 
significant fraction ($\approx17\textrm{--}40\%$) 
of intrinsically \hbox{X-ray} weak BAL quasars
in this sample. Subsequently, \nustar\ observations of the nearest
BAL quasar, Mrk~231, obtained hard X-ray spectra with sufficient 
photon statistics for spectral fitting. 
A joint \chandra\ and \nustar\ spectral analysis,
though challenging due to the substantial spectral complexity present,
suggests Compton-thin absorption 
($N_{\rm H}\approx1.2\times10^{23}$~cm$^{-2}$), making Mrk~231 intrinsically
X-ray weak by a factor of $\approx10$ \citep{Teng2014}. 
These \nustar\ and \chandra\ results
provide the first clear evidence for the existence of 
intrinsically
X-ray weak BAL quasars.

As an extension of the L13 pilot program, we obtained 
\hbox{20--35~ks} \nustar\ observations of an additional six
BAL quasars that show significant
\hbox{X-ray} weakness in the $<10$~keV band. The aim is to 
evaluate whether they show similar hard X-ray weakness to the pilot sample.
The nature of the \hbox{X-ray} weakness, whether intrinsic or due to Compton-thick 
absorption, can be assessed via stacking analyses or statistical 
arguments for the full sample of eight 
BAL quasars, including the two L13 objects.
We describe
the sample selection and \nustar\ data analysis in Sections 2 and 3.
The stacking analyses and column-density constraints are presented in
Section 4. We discuss the possibility of intrinsic
X-ray weakness for our sample in Section~5, and we summarize in Section~6. 
Throughout this paper,
we use J2000.0 coordinates and a cosmology with
$H_0=70.4$~km~s$^{-1}$~Mpc$^{-1}$, $\Omega_{\rm M}=0.272$,
and $\Omega_{\Lambda}=0.728$ \citep[e.g.,][]{Komatsu2011}.
Full names of the targets are listed in the tables while abbreviated
names are used in the text.
We quote uncertainties at a 1$\sigma$ confidence level
and upper and lower limits at 
a 90\% confidence level, unless otherwise stated.

\section{SAMPLE SELECTION} \label{sec-sample}

We selected BAL-quasar targets based on the following four criteria: 

\begin{enumerate}
\item
The targets are bona-fide BAL quasars with \ion{C}{4}~$\lambda1549$ 
absorption-trough widths $>2000$~km~s$^{-1}$.

\item
The targets are optically
bright ($m_B\lesssim16$) so that we would 
expect a significant number of 
hard X-ray photons detected with \nustar\ provided 
they have nominal underlying 
\hbox{X-ray} emission as expected from
the \hbox{$\alpha_{\rm OX}$--$L_{\rm 2500~{\textup{\AA}}}$}
relation and they are not Compton-thick.

\item
The targets are significantly \hbox{X-ray} 
weak in the \hbox{$\la10$~keV} band with
\hbox{X-ray} weakness that cannot be accounted for
by any apparent \hbox{X-ray} absorption determined using
\chandra\ or \xmm\ data.\footnote{We require 
a factor of $\gtrsim10$ times X-ray weakness at 
$\approx2$~keV by comparing 
the measured $\alpha_{\rm OX}$ parameter
to the one expected from the \hbox{$\alpha_{\rm OX}$--$L_{\rm 2500~{\textup{\AA}}}$} relation, and 
also a factor of $\gtrsim2$ times X-ray weakness in the 
observed \hbox{2--8~keV} band by measuring the $\alpha_{\rm OX,corr}$
parameter (see Section~1.1 of L13 for discussion of $\alpha_{\rm OX,corr}$).
In fact, the objects ultimately selected for our sample generally
exceed these requirements by a substantial margin (see below).}
Therefore, they are either
intrinsically X-ray weak or heavily obscured.
\nustar\ $>10$~keV observations will help to discriminate
between these two scenarios.

\item
The targets are radio quiet (radio-loudness parameter
$R<10$)\footnote{$R=f_{5~{\rm GHz}}/f_{\rm 4400~{\textup{\AA}}}$
\citep[the ratio between the 5~GHz and 4400~\AA\ flux densities
in the rest frame; e.g.,][]{Kellermann1989}.
We obtained radio flux information from the Faint Images of the Radio Sky at
Twenty-Centimeters (FIRST) survey \citep{Becker1995} 
or the
NRAO VLA Sky Survey (NVSS; \citealt{Condon1998}).}
so that their X-ray emission is not
contaminated by any jet-linked emission.

\end{enumerate}

\noindent We searched for such targets in the $z<0.5$
Palomar-Green (PG) quasar sample \citep{Schmidt1983} and in
literature reports of BAL quasars with significant \hbox{X-ray}
weakness. 
The six targets are listed in Table~1, of which
PG~0043 and PG~1001 are among the $z<0.5$ PG quasar sample, while
IRAS~07598 
\citep[e.g.,][]{Gallagher1999,Imanishi2004,Saez2012}, 
PG~0946 \citep[e.g.,][]{Mathur2000,Saez2012}, 
PG~1254 \citep[e.g.,][]{Sabra2001},
and IRAS~14026 \citep[e.g.,][]{Saez2012} are from the literature.
These BAL quasar targets have $B$-band magnitudes of
$\approx15$-16, and they are among the
optically brightest and most luminous
BAL quasars known at $z<1.3$ (Figure~1).
The more luminous PG~0946 and PG~1254 are also representative
counterparts of the luminous BAL quasars typically studied at $z\approx2$--3
\citep[e.g.,][]{Gibson2009}.

There are five BAL quasars among the $z<0.5$ PG quasar sample
\citep[see Footnote~4 of][]{Brandt2000}, four of which have now been
included in our \nustar\ BAL quasar program (PG~0043 and PG~1001
here and PG~1004 and PG~1700 in L13).\footnote{The other BAL
quasar is PG~2112+059, the X-ray weakness of which can be explained by
moderate absorption \citep[e.g.,][]{Gallagher2001,Gallagher2004}.} 
Therefore, we are sampling
a significant fraction of the most luminous BAL quasars at low redshifts.
The sample completeness
among the general population of BAL quasars is not clear
as there have been no
systematic identifications of BAL quasars at low redshifts, which
usually require {\it Hubble Space Telescope} spectra covering the key
\ion{C}{4}~$\lambda1549$ transition.

\begin{figure*}
\centerline{
\includegraphics[scale=0.5]{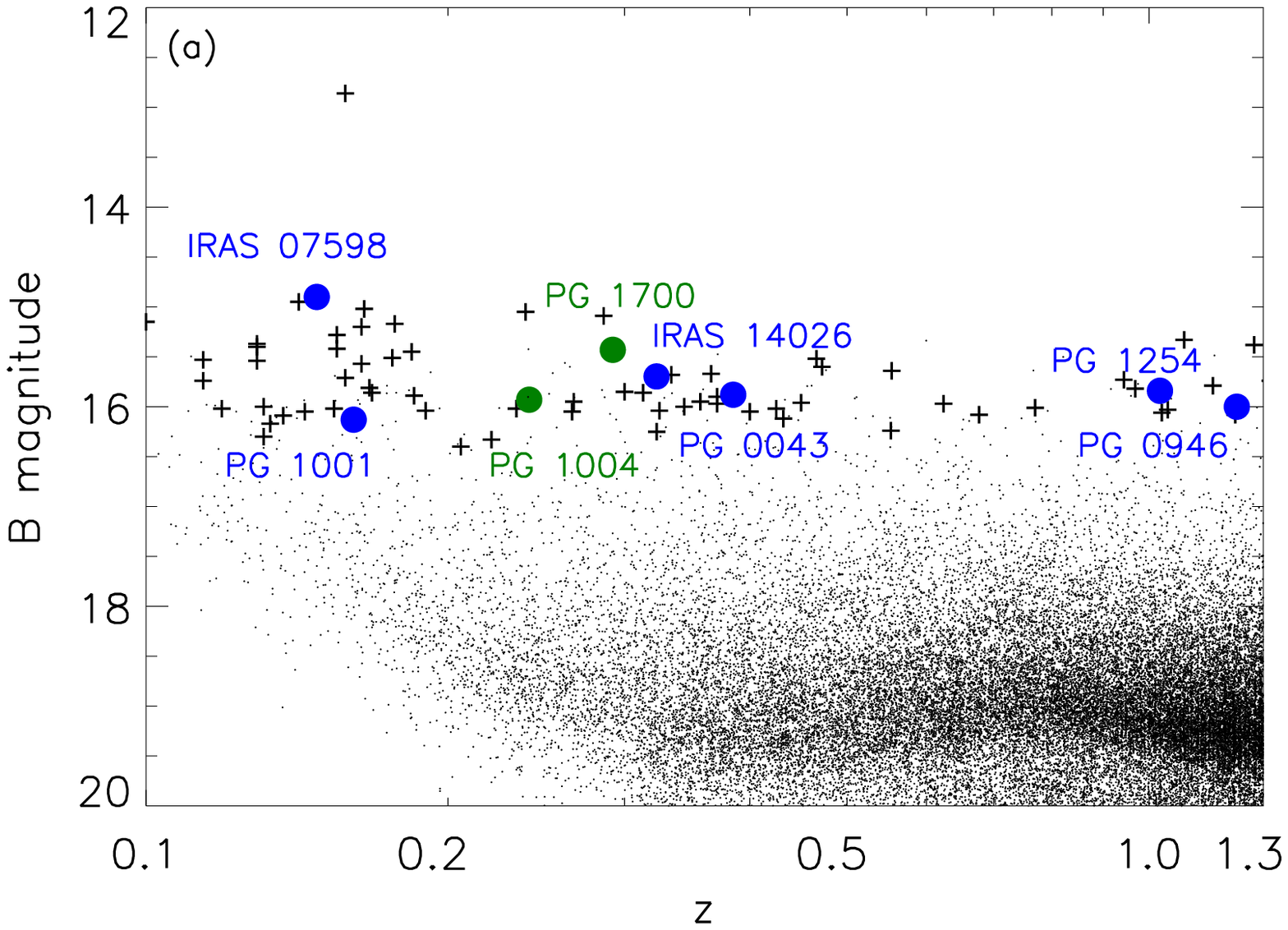}
\includegraphics[scale=0.5]{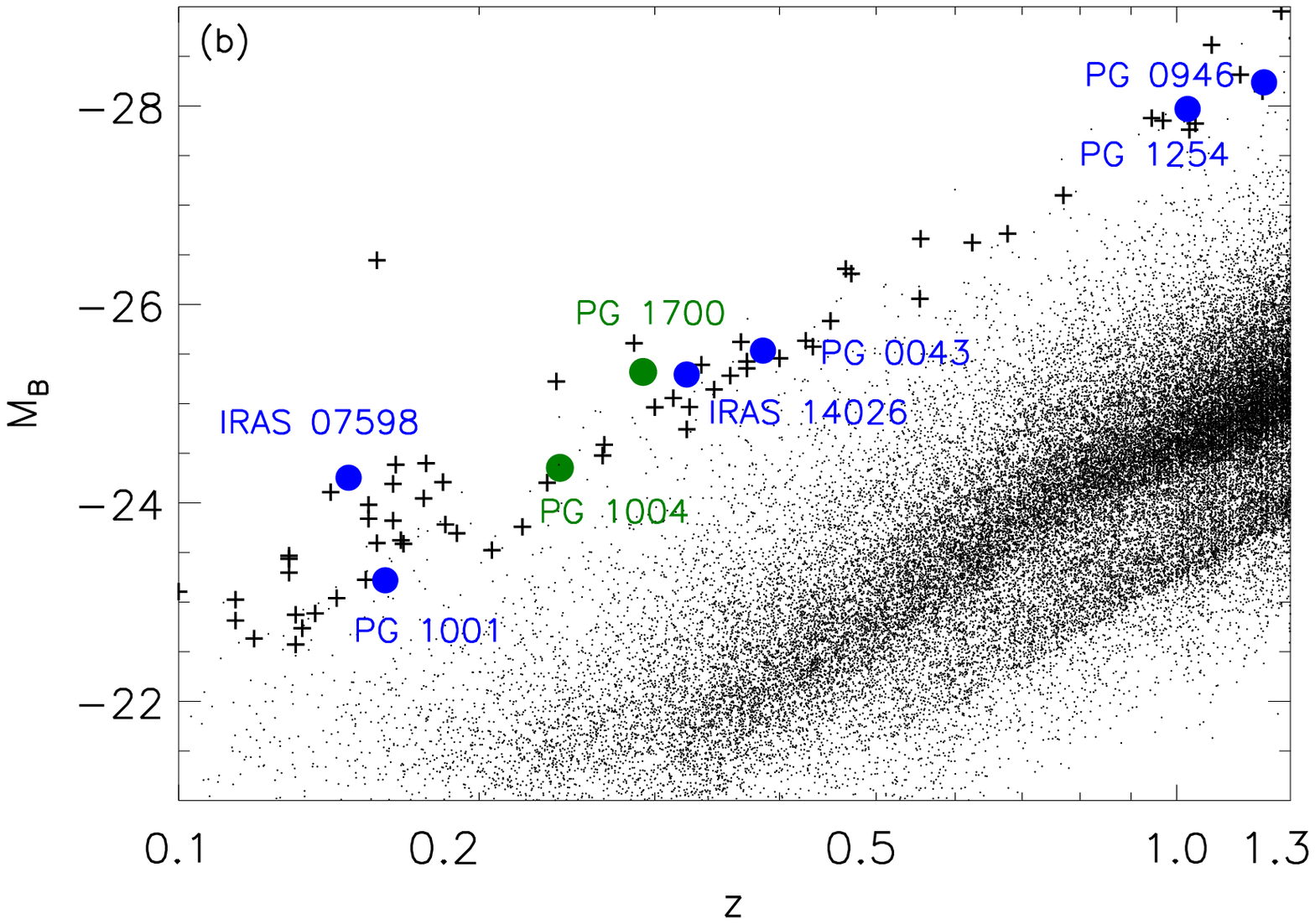}
}
\caption{
Redshift vs. (a) apparent and (b) absolute $B$-band magnitudes
for our sample objects (blue filled circles).
The green filled circles represent
the two targets in L13. The plus signs are
the bright PG quasars from \citet{Schmidt1983}; the brightest 
object is PG~1226+023 (3C 273).
The underlying black dots are objects from the SDSS DR7 quasar
catalog \citep{Schneider2010}.
The $B$-band
magnitudes of the SDSS quasars were converted from the $g$-band
magnitudes, assuming an optical power-law slope of $\alpha_{\rm o}=-0.5$
($f_\nu\propto \nu^{\alpha}$; e.g., \citealt{Vandenberk2001}). The
$K$-corrections were performed assuming the same optical power-law slope.
Our targets are among the
optically brightest and most luminous
BAL quasars known at $z<1.3$.
\label{fig-lz}}
\end{figure*}

\begin{figure}
\centerline{
\includegraphics[scale=0.5]{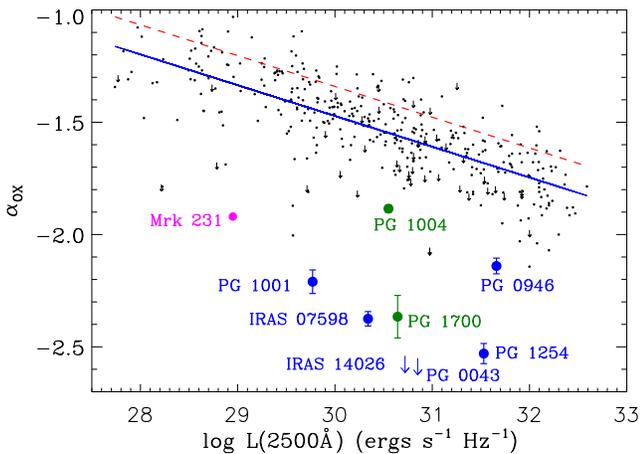}
}
\caption{
X-ray-to-optical power-law slope ($\alpha_{\rm OX}$) vs.\ 2500 \AA\
monochromatic luminosity (not corrected for intrinsic reddening) 
for the six
targeted X-ray weak BAL quasars (blue data points).
Also shown are the two objects in the pilot sample (green; L13)
and Mrk~231 (magenta; \citealt{Teng2014}).
The small black dots and downward arrows (upper limits)
are from the sample of Steffen et al. (2006) with the solid blue
line showing the $\alpha_{\rm OX}$--$L_{\rm 2500~{\textup{\AA}}}$ relation.
The dashed red line represents 
the Steffen et al. (2006) relation modified
with the excess X-ray luminosity expected for the radio loudness of
PG~1004+130 \citep{Miller2011}. 
All these BAL quasars are significantly X-ray weak
at rest-frame $\approx2$~keV.
\label{fig-aox}}
\end{figure}

All six targets have archival \chandra\ and/or \xmm\ observations.
The X-ray observations of IRAS~07598, PG~0946, PG~1001, and IRAS~14026
are summarized in \cite{Saez2012} and references therein. 
PG~1254 has a 36~ks \chandra\
observation that was reported in \citet{Sabra2001}. PG~0043 has two
$\approx30$~ks \xmm\ observations.
The source was not detected in the first observation and 
data for the second observation are not publicly
available yet.
We reprocessed the public \chandra\ and \xmm\ 
data and obtained the $\alpha_{\rm OX}$
parameters for the targets, 
with $f_{2~{\rm~keV}}$
derived from the observed 0.5--2~keV flux and 
$f_{2500~{\textup{\AA}}}$ derived by
interpolating the \hbox{optical--UV} photometric data (see Section~5.1 below).
The $\alpha_{\rm OX}$ constraints for the targets are shown in Figure~2, 
which indicate that
their rest-frame $\approx2$~keV luminosities
are 14 to $>330$ times weaker than
expected from the \citet{Steffen2006}
\hbox{$\alpha_{\rm OX}$--$L_{\rm 2500~{\textup{\AA}}}$} relation.
These $\alpha_{\rm OX}$ values have not been
corrected for any intrinsic optical/UV reddening, which would render the
values even more negative.
Note that the spread of the \citet{Steffen2006} data points 
constrains 
the expected amount of quasar variability \citep[e.g.,][]{Gibson2012},
and after accounting for the measured flux variability of our targets 
(see Section~3 below),
they are still significantly X-ray weak.
Also evident in Figure~2 is the significant $\approx2$~keV weakness
of PG~1004, PG~1700, and Mrk~231 (L13; \citealt{Teng2014}).
In addition, we verified that these targets are still
X-ray weak (by factors of $\approx2$ to $>220$) in a somewhat 
harder band
by measuring $\alpha_{\rm OX,corr}$
with the observed \hbox{2--8~keV} flux assuming 
a $\Gamma=1.8$ power-law spectrum
(see Section~1.1 of L13).
Given the significant \hbox{X-ray} weakness of these six targets, they could
be either heavily obscured or intrinsically X-ray weak.

There have been no previous tight
constraints on the \hbox{$>10$~keV} 
emission of these 
objects. They are not detected in the 70~month {\it Swift}-BAT
14--195~keV all-sky survey \citep{Baumgartner2013}, the sensitivity
of which falls short by an order of magnitude even if they have
nominal hard \hbox{X-ray} emission.
With the much higher sensitivity of \nustar,
the hard X-ray emission of these objects can be sensitively 
constrained with
relatively short exposure times.
Assuming that the targets are intrinsically X-ray normal with
an underlying 2~keV luminosity determined by the
\hbox{$\alpha_{\rm OX}$--$L_{\rm 2500~{\textup{\AA}}}$} relation,
we can estimate the \nustar\ counts yield 
following
the approach in Section~4.1.1 of L13 using the {\sc MYTorus} model \citep{Murphy2009}.
Provided that the sources are Compton-thin
($N_{\rm H}<1.5\times10^{24}$~cm$^{-2}$), we would expect
significant detections
of PG~0043, IRAS~07598, PG~1001, and IRAS~14026 in the \hbox{8--24~keV} band 
(more than $80$--260 counts where these values are derived for
$N_{\rm H}=1.5\times10^{24}$~cm$^{-2}$) with 20~ks \nustar\
observations. Longer exposures are
required for the more distant targets PG~0946 (35~ks) and PG~1254
(30~ks) for a 8--24~keV detection with more than $30$ counts. 
Fewer counts are expected, of course, 
if the targets are either intrinsically X-ray
weak or have Compton-thick column densities substantially exceeding
$1.5\times10^{24}$~cm$^{-2}$.

\section{\nustar\ OBSERVATIONS AND DATA ANALYSIS} \label{sec-data}
The details of the \nustar\
observations are listed in Table~\ref{tbl-obs}.
We processed the level 1 data using 
the \nustar\ Data Analysis Software (NuSTARDAS)
v.1.2.0 with \nustar\ CALDB 20131007, and produced
cleaned calibrated event files (level 2 data) 
using the {\sc nupipeline} script
for both
focal plane modules (FPMs, including FPMA and FPMB; \citealt{Harrison2013}).
For each target in each FPM,
we created \hbox{X-ray} images in four bands:
\hbox{3--24~keV}, \hbox{3--8~keV}, \hbox{8--24~keV},
and \hbox{24--79~keV} using
the \chandra\ Interactive Analysis
of Observations (CIAO)\footnote{See
http://cxc.harvard.edu/ciao/ for details on CIAO.}
v4.5 tool {\sc dmcopy}.
These bands are being adopted as standard photometric bands
in current \nustar\ studies \citep[e.g.,][]{Alexander2013,Lansbury2014}, but
they are slightly different from those used in the early L13 pilot study 
(e.g., \hbox{4--20~keV}, \hbox{4--10~keV}, and \hbox{10--20~keV} bands).
We have verified that alternative choices of photometric bands
yield consistent results.
 
We searched for sources in each of the images
using the CIAO tool {\sc wavdetect} \citep{Freeman2002}
with a false-positive probability
threshold of 10$^{-5}$ and wavelet scales of 2, 2.83, 4, 5.66, 8, 11.31, 
and 16 pixels (the pixel size is 2.46\arcsec).
PG~0946, PG~1001, and PG~1254 are detected in at least one band
in each FPM. 
The minimum positional offsets (Table~\ref{tbl-obs}) 
between the optical and 3--24~keV positions are within expectations 
for faint sources.
The chance of getting any spurious detections by {\sc wavdetect}
at these known source positions is negligible.
The other three targets, PG~0043, IRAS~07598, and IRAS~14026, 
are not detected; we also verified 
the non-detections via visual inspection of the 
smoothed images. None of the targets
is detected in the \hbox{24--79~keV} band, and 
the constraints from the non-detections
in this band are not as tight as those in the other bands. 
Stacking in this band does not yield any useful constraints either.
Therefore, we do not include the 24--79~keV 
band in the following discussion.

We performed aperture photometry for each target in the 
three standard bands above. 
Total (source plus background) counts 
were extracted within a 35\arcsec-radius circular 
aperture, centered on the \hbox{3--24~keV}
position (for detected targets)
or optical position (for undetected targets).
This aperture
approximates the
$63.9\%$ encircled-energy fraction contour of the point
spread function, and we have verified that 
different choices of the source-extraction region yield consistent results.
Background counts were extracted from a simulated background map
created using the {\sc nuskybgd} script \citep{Wik2014}; these are
consistent with those estimated from 
annular or circular off-source regions.
We followed the binomial no-source probability ($P_{\rm B}$)
approach in L13 to determine the source detection significance 
in each \nustar\ band.
If the $P_{\rm B}$ value
is smaller than 0.01 ($\approx2.6\sigma$),
we considered the source detected and calculated
the 1$\sigma$ errors on the net counts \citep{Gehrels1986}.
If the $P_{\rm B}$ value is larger than 0.01, we considered the
source undetected and derived
an upper limit on the source counts
using the Bayesian approach of \citet{Kraft1991}.
The aperture-corrected source counts and upper limits
for our targets are listed in Table~2.
Measurements for FPMA and FPMB are consistent within the uncertainties.
In the \hbox{8--24~keV} band, none of the targets is detected
except PG~0946 in FPMA with a $23$-count detection;
the source counts are below our expectations for the Compton-thin
scenario (Section~2), and thus all targets are hard X-ray weak, similar
to the pilot sample in L13.

Since a relatively large aperture with a 35\arcsec\ radius 
was used in photometry extraction,
we investigated whether there are any contaminating sources nearby.
We inspected the \chandra\ or \xmm\ images for our targets, and confirmed
that there is no neighboring source within 50\arcsec\ of the targets except
PG~0043. There is one source 25\arcsec\ away from PG~0043 that was
reported in \citet{Ballo2008}. The \xmm\ data show a soft spectrum
($\Gamma=1.86$) with a 2--10~keV flux of $1.8\times10^{-14}$~\flux.
We estimated that the contamination from this source in our aperture
is negligible ($\approx0.8$ counts in the 3--8~keV band and $\approx0.5$
counts in the 8--24~keV band). This source should not affect our 
stacking results below (Section~4.1) either.

Following L13, we derived a
3--24 keV effective power-law photon index ($\Gamma_{\rm eff}$) from the 
band ratio between the 8--24~keV and \hbox{3--8~keV} bands, calibrated
using the \nustar\ spectral response files extracted at the source 
location and assuming a power-law spectrum with the
Galactic absorption column density \citep{Dickey1990}. 
The uncertainties of (or limits on) 
the band ratios (and subsequently 
$\Gamma_{\rm eff}$) were
derived using the
Bayesian code {\sc behr} \citep{Park2006}.
For sources undetected in both the 8--24~keV and \hbox{3--8~keV} bands,
$\Gamma_{\rm eff}=1.8$ was adopted.
The $\Gamma_{\rm eff}$ values are listed in Table~2, which do not individually
provide tight constraints on whether the sources have hard (indicative
of absorption) or soft spectra, due to the non-detections or large 
uncertainties.
The source fluxes and luminosities, listed in Table~2, 
were converted from the count rates
and $\Gamma_{\rm eff}$, calibrated with the spectral response files.

We compared the \nustar\ 3--8 keV flux measurements to previous 
\chandra\ and \xmm\ data. Long-term flux variability is observed in the 
three detected targets.
The \hbox{3--8~keV} flux of PG~0946 has dropped by a factor of $5.4\pm2.8$ 
between the
2010 \chandra\ observation \citep{Saez2012} and the \nustar\ observation. 
The \nustar\ flux of PG~1001 is consistent with that measured in the 2003 
\xmm\ observation, and it is $4.9\pm2.9$ times higher than the flux in the 
2010 \chandra\ observation \citep{Saez2012}. PG~1254 is $2.7\pm1.4$ times 
brighter in the \nustar\ observation compared to the 2000 
\chandra\ observation \citep{Sabra2001}. 
We note that after accounting for their flux increases, 
PG~1001 and PG~1254 
are still 21 and 58 times X-ray weak at $\approx2$~keV (e.g., in Figure~2),
respectively.
Similar flux variability has also 
been noted in PG~1004 in the pilot sample 
(\citealt{Miller2006}; L13) 
and several other BAL quasars \citep[e.g.,][]{Gallagher2004,Saez2012}. 
For the other three 
undetected targets, the flux upper limits on
IRAS~07598 and IRAS~14026 are consistent with previous \chandra\
and/or \xmm\ flux measurements, and PG~0043 is not detected by \xmm\ either.
For IRAS~07598 and IRAS~14026, combining the 3--8~keV fluxes from 
\chandra\ or \xmm\ and the \hbox{8--24~keV} flux upper limits from \nustar\
does not provide useful constraints on $\Gamma_{\rm eff}$.
It is not useful to combine the lower energy data with the \nustar\ data
for the three detected targets due to the observed variability.

\section{RESULTS}

\subsection{Stacking and Joint Spectral Analyses}
Since our targets are only weakly detected or undetected by \nustar,
we cannot study the
nature of their hard \hbox{X-ray} weakness via individual spectral analysis.
Instead, we performed stacking and joint spectral 
analyses to probe the average spectral 
properties of the sample. 
First, we stacked the FPMA and FPMB data for each object
individually. The three undetected targets are still not detected.
PG~0946 and PG~1254 are detected in the 8--24~keV band, allowing better
constraints on their effective photon
indices. For PG~0946, we obtained a
$\Gamma_{\rm eff}$ of $1.2_{-0.6}^{+0.7}$, consistent with its FPMA measurement.
For PG~1254, $\Gamma_{\rm eff}$ is $1.5_{-0.6}^{+0.8}$.

Stacking of the full sample of six objects
in both FPMA and FPMB 
yields significant detections in both the 3--8~keV and 8--24~keV bands,
with $164.5_{-29.4}^{+31.0}$ and $102.3_{-28.3}^{+29.9}$ counts, respectively.
The band ratio to $\Gamma_{\rm eff}$ conversion factors 
vary slightly between different
sources, and thus we adopted the average value to convert the 
8--24~keV to 3--8~keV band ratio of the stacked source to
an effective photon index, which is $\Gamma_{\rm eff}=1.8_{-0.4}^{+0.5}$.
This $\Gamma_{\rm eff}$ represents the weighted
average of the individual effective photon
indices; the weight varies (by factors of a few) 
between sources 
due to their different fluxes, exposure times, and rest-frame bands probed.
For the subsample of the
three detected targets, the stacked counts in the 3--8~keV and 8--24~keV
bands are $115.7_{-21.8}^{+23.5}$ and $92.0_{-22.1}^{+23.7}$, respectively,
and the effective photon index is $\Gamma_{\rm eff}=1.5_{-0.4}^{+0.4}$;
for the undetected subsample, the stacked source is only weakly
detected in the 3--8 keV band, and the stacked counts in the 3--8~keV and 
8--24~keV bands are
$48.8_{-19.7}^{+21.3}$ and $<38.1$, respectively, with $\Gamma_{\rm eff}>0.8$. 
Given the $\Gamma_{\rm eff}$ values for the stacking of the
full sample and the detected subsample, the stacked source for the 
undetected subsample is likely soft, e.g., the lower limit on 
$\Gamma_{\rm eff}$ at a less conservative 1$\sigma$ confidence level is 
$\Gamma_{\rm eff}>1.3$.

The stacked 3--8~keV counts should have negligible contribution from
host-galaxy X-ray emission; the contribution is only $\approx3\%$
for host galaxies with a high \hbox{X-ray} luminosity 
of $10^{42}$~\lum, which corresponds to a star-formation rate
of $\approx620~M_{\sun}$~yr$^{-1}$ \citep[e.g.,][]{Lehmer2010}.
Therefore, the stacked signals are dominated by the nuclear sources.
The soft stacked effective photon indices
for the full sample ($1.8_{-0.4}^{+0.5}$) and the detected 
subsample ($1.5_{-0.4}^{+0.4}$)
suggest that the targets on average are not
absorbed by Compton-thick material, which would generally result in 
a flat ($\Gamma_{\rm eff}\approx0.5$ with a range of $\approx0\textrm{--}1$) spectrum
from 3--24~keV.
Moreover, 
the soft stacked signal for the full sample 
cannot be dominated by one single object given their 
individual count contributions in the 3--8~keV band (three sources are not
detected and the other three are weakly detected), indicating that 
at least two objects are responsible for the soft stacked $\Gamma_{\rm eff}$.
Therefore, at least 33\% of the sample objects likely 
have soft effective photon indices ($\Gamma_{\rm eff}\ga1.8$) 
and are likely not Compton-thick.

We also jointly fitted the \nustar\ spectra of the three detected targets,
PG~0946, PG~1001, and PG~1254, with XSPEC v.12.8.1g \citep{Arnaud1996}.
We used the {\sc nuproducts} script of NuSTARDAS to extract source spectra
within the same 35\arcsec-radius circular
apertures as in the photometry extraction, and local background spectra
within annular regions with inner and outer radii of 
120\arcsec\ and 240\arcsec, respectively.
The 3--24~keV spectra for the three targets in both FPMA and FPMB
were fitted jointly with a simple power-law model using the $C$ 
statistic (cstat) in XSPEC,\footnote{The $W$
statistic was actually used in the presence of background
spectra; see
http://heasarc.gsfc.nasa.gov/docs/xanadu/xspec/manual/XSappendixStatistics.html.} allowing each target to have its own
redshift and Galactic column density. The best-fit photon index is 
$\Gamma=1.55_{-0.29}^{+0.30}$ ($C=290$ for 335 degrees of 
freedom), consistent with the stacked
effective photon index for this subsample. The limited photon statistics 
of the spectra
do not allow for any useful constraint on Fe K$\alpha$ line emission at
rest-frame 6.4--6.97~keV; a strong
Fe~K$\alpha$ line with an equivalent width of order 1--2~keV
might be expected if the continuum is reflection dominated
\citep[e.g.,][]{Ghisellini1994,Matt1996,Gandhi2014}.
Previous \chandra\ or \xmm\ observations do not provide useful 
constraints on the 
Fe K$\alpha$ line either \citep[e.g.,][]{Imanishi2004}.

\subsection{Indirect Absorption Column-Density Constraints}

We adopted the same 
approach described in Section~4.1.1 of L13 to constrain the
absorption column densities indirectly for the six targets, 
under the assumption that the observed 
hard \hbox{X-ray} weakness is attributed entirely to absorption and they have
nominal underlying X-ray emission as
determined 
from the \hbox{$\alpha_{\rm OX}$--$L_{\rm 2500~{\textup{\AA}}}$} relation.
Briefly, an absorption column density was derived by comparing the 
observed flux to the expected one derived from the expected $\alpha_{\rm OX}$
assuming a power-law \hbox{X-ray} spectrum with $\Gamma=1.8$. 
The {\sc MYTorus} XSPEC model, including both the 
transmitted and scattered spectral components,
was used to calibrate the relation between 
$N_{\rm H}$ and this X-ray weakness.\footnote{The physical properties of the
shielding gas in BAL quasars are poorly understood \citep[e.g.,][]{Proga2004}.
The parameterization of
the {\sc MYTorus} model cannot fully reproduce the
complex absorption environments of our targets, but we consider
it the best available approximation for the purpose of deriving
basic column-density constraints.}
We assumed 
a \hbox{half-opening} angle of 60\degr\
(corresponding to a torus
covering factor of 0.5) and an inclination
angle of 80\degr\ in the {\sc MYTorus} model. The $N_{\rm H}$ dependence
on the assumed \hbox{half-opening} angle is relatively small, as illustrated
in Figures~7 and 8 
of L13 ($\approx20\%$ smaller for a \hbox{half-opening} angle of 
37\degr). Large inclination angles are generally expected for BAL
quasars in the disk-wind scenario; for inclination angles smaller than 80\degr,
we would derive larger $N_{\rm H}$ values by factors of up to $\approx3$
(Figures~7 and 8 
of L13). 

We derived column-density constraints using the 
\hbox{3--24~keV}, \hbox{3--8~keV}, and \hbox{8--24~keV}
\nustar\ fluxes or flux upper limits. The constraints in the 
\hbox{3--24~keV} and \hbox{8--24~keV} bands are comparable, and
they are significantly tighter than those in the softer \hbox{3--8~keV}
band, as one would expect given the higher rest-frame energies utilized
(also see Figures~7 and 8
of L13). We list in Table~2 the factors of \hbox{3--24~keV} weakness and 
the \hbox{3--24~keV} $N_{\rm H}$ constraints.
The $N_{\rm H}$ uncertainty accounts for 
the measured flux uncertainty and the spread of the intrinsic $\alpha_{\rm OX}$
value which
was assumed to follow a Gaussian distribution with its 1$\sigma$ uncertainty
from Table~5 of \citet{Steffen2006}. For the undetected targets, the
lower limits on $N_{\rm H}$ were obtained from $N_{\rm H}$
probability distributions that were 
derived from the $\alpha_{\rm OX}$ Gaussian
distributions and the probability distributions of the 3--24~keV 
fluxes.\footnote{The probability distributions of the 3--24~keV 
fluxes are from the probability distributions of the 3--24~keV counts,
which were derived during our computation of the band ratios using 
the {\sc behr} code.}
The column density constraints indicate
that Compton-thick absorption is required for all six targets to 
produce the observed hard X-ray weakness if they have
nominal intrinsic X-ray emission, as expected from our experimental design.

We note that these column density constraints were derived from the 
observed hard \hbox{X-ray} weakness, independent of  
the spectral-shape constraints above. In fact, Compton-thick absorption
is likely inconsistent with a soft $\Gamma_{\rm eff}$, as 
discussed in more detail
below (Section~5.2).

\section{DISCUSSION}

\subsection{Multiwavelength Properties}
\begin{figure*}
\centerline{
\includegraphics[scale=0.5]{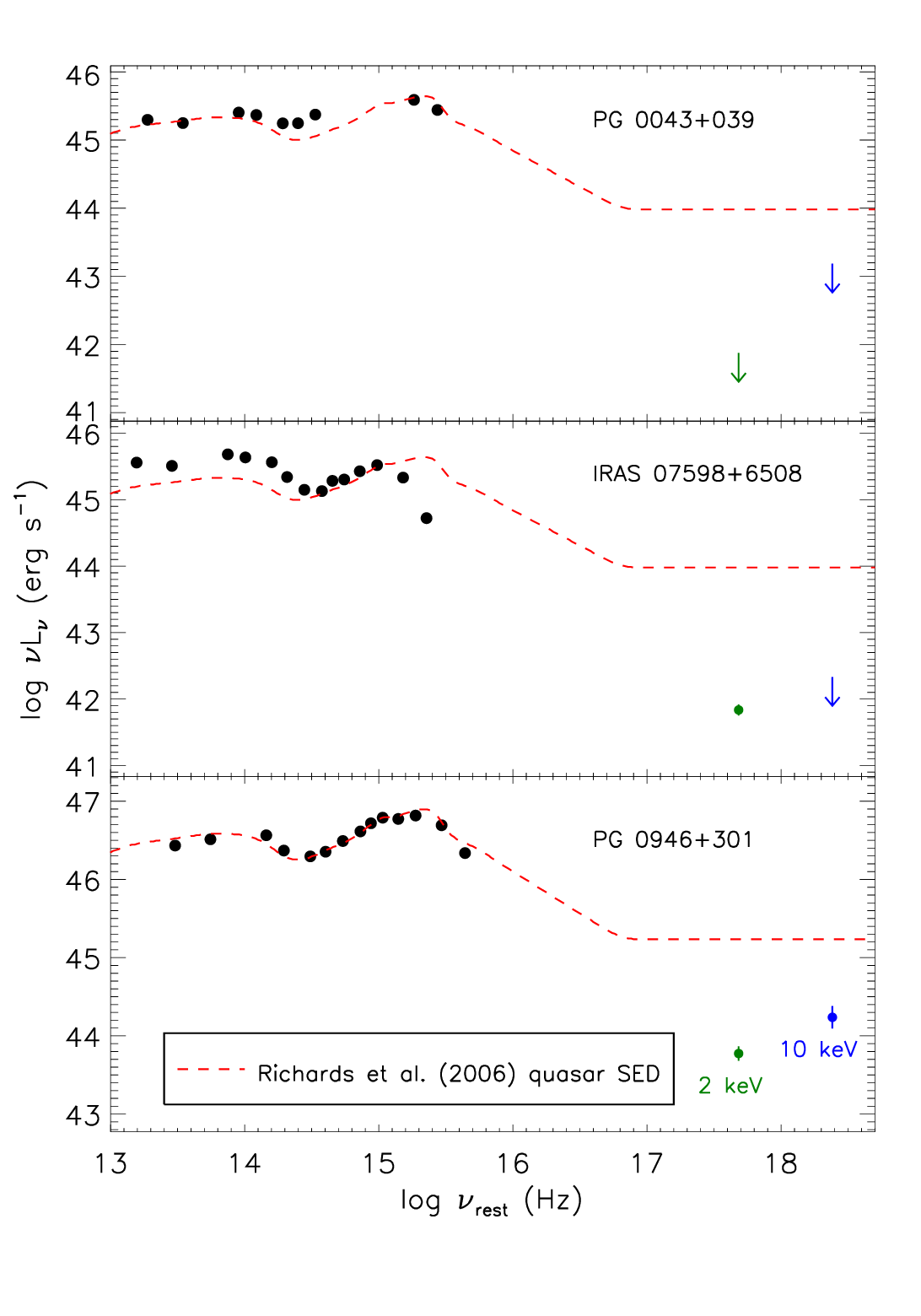}
\includegraphics[scale=0.5]{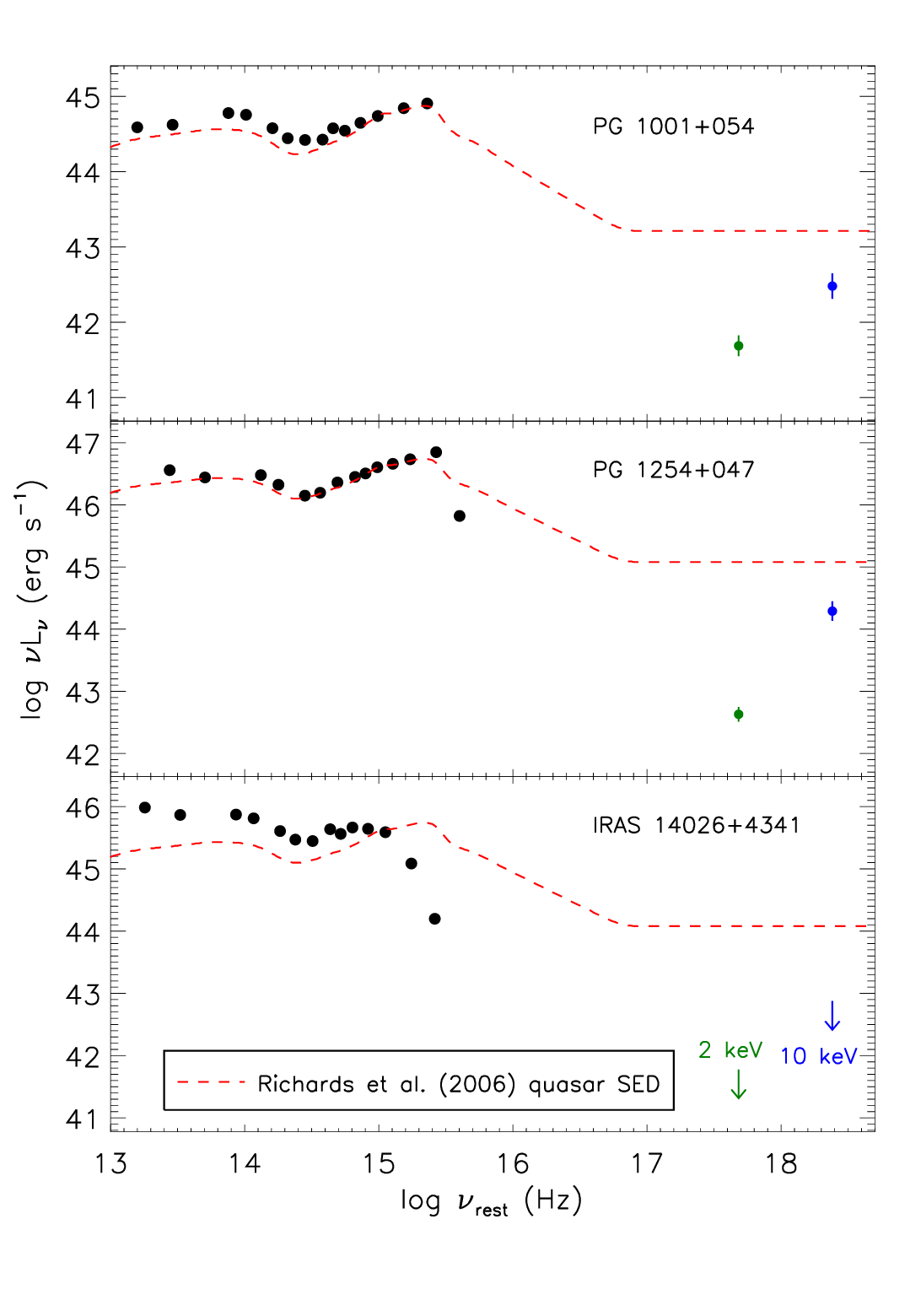}
}
\caption{IR through \hbox{X-ray} SEDs of the six targets
in the rest frame. The IR--UV data (black points) are from the {\it WISE},
2MASS, SDSS, and/or {\it GALEX} catalogs. The 2~keV data (green points
and arrows) are 
from previous \chandra\ or \xmm\ observations; note that 
IRAS~14026 was detected in the 2--8~keV band by \chandra, but not
in the 0.5--2~keV band. The 10~keV data (blue points and arrows) 
were derived from the \nustar\ 3--24 keV fluxes or 
flux upper limits averaged over FPMA and FPMB assuming $\Gamma_{\rm eff}=1.2$
for PG~0946 and $\Gamma_{\rm eff}=1.8$ for the rest of the objects.
The error bars for most of the data points are smaller than 
or comparable to the symbol size
and are thus not visible.
The SED data were not observed simultaneously and may be affected by
variability (e.g., see the X-ray variability in Section 3).
The red dashed curve shows the composite quasar SED of 
\citet{Richards2006}
normalized to the 3000~\AA\ luminosity.
\label{fig-sed}}
\end{figure*}

As in L13, we constructed infrared (IR) to X-ray
spectral energy distributions (SEDs) for the targets, using photometric
data from the {\it Wide-field Infrared Survey Explorer}
({\it WISE}; \citealt{Wright2010}), Two Micron All Sky Survey (2MASS;
\citealt{Skrutskie2006}), SDSS, and/or {\it Galaxy Evolution Explorer}
({\it GALEX}; \citealt{Martin2005}) catalogs.
The rest-frame SEDs are shown in Figure~\ref{fig-sed}.
The optical and UV data have
been corrected for Galactic extinction following the dereddening
approach of \citet{Cardelli1989} and \citet{Odonnell1994}.
Besides the strong intrinsic reddening in 
IRAS~07598 and IRAS~14026
(also see \citealt{Jiang2013} for the reddening in IRAS~14026)
and the significant X-ray weakness,
these targets have typical radio-quiet quasar SEDs 
\citep[e.g.,][]{Richards2006}, similar to PG~1700 in the pilot sample.

Depending upon whether there are BALs
from ions at low-ionization states such as
\ion{Mg}{2} or \ion{Al}{3},
BAL quasars are classified as low-ionization BAL (LoBAL)
or high-ionization BAL (HiBAL) quasars 
\citep[e.g.,][]{Weymann1991,Sprayberry1992}.
LoBAL quasars constitute a minority ($\approx10\%$)
of BAL quasars, and they often show signs
of dust reddening and are \hbox{X-ray} weaker than HiBAL quasars
\citep[e.g.,][]{Green2001,Gallagher2006,Gibson2009}.
Among our six targets here, IRAS~07598 and IRAS~14026
are LoBAL quasars \citep[e.g.,][]{Hines1995,Hines2001}, while
the other objects are HiBAL quasars 
\citep[e.g.,][]{Turnshek1994,Arav2001}.
The two LoBAL quasars indeed show significant dust reddening 
and soft and hard \hbox{X-ray} weakness (Figure~3 and Table~2).

It has been suggested that significant X-ray weakness may be associated
with super-Eddington accretion \citep[e.g.,][]{Lusso2010}.
PG~1004 and PG~1700 in the pilot sample appear to have sub-Eddington accretion
(Eddington ratios $\approx0.09$ and $\approx0.41$; L13), 
while Mrk~231 indeed appears to be
a super-Eddington source with an Eddington ratio of $\approx5$ \citep{Teng2014}.
We estimated bolometric luminosities for our targets
from the \citet{Richards2006} composite quasar SED normalized
to their 3000~\AA\ luminosities (not strongly affected by 
intrinsic reddening), and collected their
single-epoch virial black hole (BH) masses
from the literature \citep{Hao2005,Shen2011}.
These data are listed in Table~1. The derived Eddington ratios
are in the range of 0.13 to 0.62, all in the sub-Eddington regime. 
However, there are significant
uncertainties associated with the estimated BH masses
\citep[$>0.3$ dex; e.g.,][]{Shen2012} and bolometric luminosities, 
and thus it is difficult to assess whether the significant \hbox{X-ray}
weakness is related to super-Eddington accretion in these cases.

The intrinsically X-ray weak quasar PHL~1811
has unusually weak and blueshifted high-ionization lines (e.g.,
the equivalent width of its \ion{C}{4}~$\lambda1549$ emission
line is 6.6~\AA, much smaller than the average value of 
$\approx30$~\AA\ for SDSS quasars; \citealt{Leighly2007b}), which may be due to
the lack of high-energy ionizing continuum photons.
A recent study
of IRAS~14026 \citep{Jiang2013} suggests that it is probably a PHL 1811
analog given the weak \ion{C}{4} and \ion{C}{3}] line emission, although
no quantitative measurement of the line strength has been given.
\citet{Wu2011} proposed
a simple unification model (e.g., Figure 9 of \citealt{Wu2011}) 
where PHL 1811 analogs and BAL quasars
have similar inner structures but the lines of sight to PHL 1811 analogs
do not intercept the UV-absorbing disk wind.
Confirmation of such a
connection would facilitate
our understanding of the nature of their \hbox{X-ray} weakness.
We visually examined the UV spectra for the other targets, the pilot sample,
and also Mrk~231 
\citep[e.g.,][]{Hines1995,Hamann1998,Arav1999,Brandt2000,Gallagher2002b}.
The \ion{C}{4} emission lines appear strong in IRAS~07598, PG~0946, 
PG~1001, PG~1254, and Mrk~231, while they are either weak or are 
contaminated by the broad absorption features in PG~0043, PG~1004, and PG~1700. 
Therefore, more than half of our targets do not 
show weak \ion{C}{4} line emission as in PHL~1811.
It is possible that 
intrinsically \hbox{X-ray} weak quasars may exhibit
normal optical--UV emission lines if 
the accretion still produces 
sufficient EUV (perhaps $\la0.5$~keV) radiation
to ionize the broad line regions, e.g., photons with energies exceeding
48~eV are required to ionize \ion{C}{3} and produce the \ion{C}{4} emission 
line.
A connection between BAL quasars and PHL 1811 analogs thus cannot
be excluded.

\subsection{Intrinsic X-ray Weakness in the Sample}

As discussed extensively in L13, the hard \hbox{X-ray} weakness 
of these BAL quasars observed by 
\nustar\ can, in principle,  
be explained in either the Compton-thick absorption or 
intrinsic \hbox{X-ray} weakness scenarios. Without high photon counts,
it is not feasible to 
constrain their nature individually via spectral analyses as was done 
for the case of the local object Mrk~231. However, we have already established 
that there is apparently
a population of intrinsically \hbox{X-ray} weak BAL quasars 
(L13; \citealt{Teng2014}), and we present below some evidence
that at least some of our targets here are also intrinsically X-ray weak:

\begin{enumerate}
\item
The soft stacked effective photon index 
($1.8_{-0.4}^{+0.5}$) for the sample (Section~4.1) 
argues against Compton-thick absorption in general.\footnote{For the 
two targets in L13, PG~1004 has a soft effective photon index
($1.8\pm0.5$) but it could be dominated by jet emission, and
PG~1700 has a hard effective photon index but with a large uncertainty
($0.5\pm0.7$). The nature of their hard \hbox{X-ray} weakness is not 
clear.} This is
similar to the independent 
\chandra\ stacking results in L13 where we also found a 
relatively
soft stacked signal for a subsample of high-redshift BAL quasars, which
suggests that the stacked source is not Compton-thick but is 
intrinsically X-ray weak. A soft 3--24~keV effective photon index
is also consistent with the lower energy \chandra\ or \xmm\ spectral 
fitting results (based on $\approx45\textrm{--}320$ X-ray photons) for 
IRAS~07598, PG~0946,
PG~1001, and PG~1254 where
no or only moderate ($\la10^{23}$~cm$^{-2}$) absorption was found
\citep[e.g.,][]{Sabra2001,Imanishi2004,Schartel2005,Saez2012}.\footnote{For the remaining two objects, 
PG~0043 is not detected by \xmm\ and IRAS~14026 is weakly detected by
\chandra\, and thus spectral analysis is not possible.} 

\item 
All six targets here and the two in the L13 pilot sample 
are significantly X-ray weak in the \nustar\ bands, which
requires Compton-thick obscuration in the absorption scenario.
If these BAL quasars with significant X-ray weakness represent 
an extension of the normal BAL-quasar population to higher column densities
of $10^{23.5}$--$10^{25}$~~cm$^{-2}$,
then a first order expectation would be that they might consist of both
Compton-thin and Compton-thick objects, similar to the overall 
AGN population. In this case,
the chance of observing a sample of eight objects that are
100\% Compton-thick is likely small. A natural explanation is that 
this is not a pure Compton-thick sample and 
at least some of the targets are intrinsically X-ray weak.

In Figure~4, we show the $N_{\rm H}$ distributions for our 
\nustar\ BAL-quasar sample (eight objects in total)\footnote{The 
radio-loud nature of PG~1004 in L13
does not affect our analysis here.
If its observed X-ray emission 
has a significant jet-linked contribution, 
the estimated $N_{\rm H}$ value would be larger (see Section~4.1.1 of L13).}
and those BAL quasars collected from the literature 
\citep{Gallagher2002a,Giustini2008,Fan2009,Streblyanska2010,Morabito2014}.\footnote{These 
are the X-ray studies of large samples of BAL quasars including
$N_{\rm H}$ constraints in the literature.
We caution that these data might not represent the real $N_{\rm H}$
distribution as the sample is not complete and the 
$N_{\rm H}$ constraints were obtained via different approaches.}
The combined distribution appears disjoint and perhaps 
bimodal, missing objects that
are heavily obscured but Compton-thin, while
the \nustar\ sample stands out in the Compton-thick regime forming 
an apparently distinct peak. 
The $N_{\rm H}$ distribution for typical low-redshift
AGNs does not show such a bimodality (e.g., Figure~4 of \citealt{Ueda2014}).
There is no obvious reason why BAL quasars
would avoid the $N_{\rm H}\approx(5\textrm{--}20)\times10^{23}$~cm$^{-2}$ 
regime yet not higher $N_{\rm H}$ values, 
although we caution that our \nustar\
sample size is still relatively small and 
the apparent bimodality might be caused by small number statistics
(moreover, we note that the shielding gas responsible for the X-ray absorption
in BAL quasars is different from the dusty torus in typical obscured AGNs,
and its nature is poorly understood).
Thus, as a complement to the spectral-shape argument above,
Figure~4 provides suggestive additional
evidence that some of the \nustar\ objects 
are not absorbed by Compton-thick material but are
intrinsically X-ray weak.
 
\end{enumerate}

In obscured AGNs, there could be an additional soft \hbox{X-ray} 
continuum component
arising from electron scattering of the intrinsic continuum in an ionized
medium surrounding the central engine on a larger scale than the 
X-ray absorber. The scattering zone has a very small column density
so that the scattered continuum has approximately the same shape as the
intrinsic continuum.\footnote{The location and physical properties of the
scattering medium in BAL quasars are uncertain. The shielding gas
itself might produce a relatively soft scattered continuum if it
is sufficiently highly ionized \citep[e.g.,][]{Proga2004,Ross2005,Garcia2010}.} 
Because of this, the scattered fraction is expected
to be small, usually a few percent ($\approx5\%$) or less 
(e.g., 
\citealt{Turner1997,Ogle1999,Ueda2007,Young2007} and references therein). 
In principle, this ionized scattered component could perhaps dominate
over the Compton-reflected component in the 
observed 3--24~keV emission of a Compton-thick AGN if
the column density is sufficiently high
($N_{\rm H}>10^{25}$~cm$^{-2}$).
In this case, the observed X-ray spectrum would
appear soft and also be hard \hbox{X-ray} 
weak by a factor of $\approx20$ or more.
This scenario could perhaps explain the soft stacked signal of
our \nustar\ sample here without invoking intrinsic X-ray weakness. However,
such a case would arguably be even more extraordinary than the discovery of 
intrinsically X-ray weak AGNs, as so far no 
compelling example of a clearly Compton-thick AGN with a soft 
$\approx3\textrm{--}24$~keV continuum 
has been found (T. Yaqoob 2014, private communication). Furthermore,
our X-ray variability detections for some objects (see Section 3)
would constrain the size of any scattering medium.
Therefore, we admit this possibility here 
but do not consider it likely.

Assuming that the observed hard X-ray weakness is entirely intrinsic with
no absorption, these targets are intrinsically \hbox{X-ray} weak 
from 3--24~keV
by the factors 
given in Table~2 ($\approx4$ to $>25$).
If there is also
absorption present that affects the observed \hbox{3--24 keV} flux,
similar to the case of Mrk~231, the factors of 
intrinsic X-ray weakness would be smaller.
For comparison, Mrk~231 is intrinsically X-ray weak by a factor of
$\approx10$ after the absorption correction \citep{Teng2014},
and PG~1004 and PG~1700 in L13 are X-ray weak by about the same factor
in the intrinsic X-ray weakness scenario.
The underlying physics responsible
for intrinsic X-ray weakness is still unclear; some sort of
coronal-quenching mechanism might be relevant for BAL quasars (Section
4.2.1 of L13 and references therein).

There are two LoBAL and four HiBAL quasars in our sample (Section 5.1). 
We checked the stacked signals for these two groups of sources separately.
For LoBAL quasars (IRAS~07598 and IRAS~14026), 
the stacked source is only weakly detected in the \hbox{3--8~keV} band, and the 
lower limit on the effective photon index is 1.0 
($>1.4$ at a 1$\sigma$ confidence level). For HiBAL quasars,
the stacked effective photon index is $1.5_{-0.4}^{+0.4}$. 
Considering that the full sample
has a soft effective photon index ($1.8_{-0.4}^{+0.5}$), the LoBAL quasars
likely have soft photon indices.
It thus appears likely that 
both groups contain intrinsically X-ray weak quasars.

All three detected targets show significant flux variability in
the 3--8~keV band when compared to earlier observations
(Section 3). 
Usually we would not expect such significant variability in 
a Compton-thick AGN where the \hbox{3--8~keV} spectrum
is likely dominated by a Compton-reflected component, as variability
would be washed out during reflection over an extended 
region. However, for our BAL-quasar targets here, 
the X-ray absorber (shielding gas) is located
on a significantly smaller physical scale than the torus in
typical obscured AGNs ($\approx10^{16}$--10$^{17}$~cm vs.\ parsec scale)
and the variability observed is
on multi-year timescales. It is possible that 
a reflection-dominated spectrum could show 
long-term variability \citep[e.g.,][]{Matt2004},
and the scenario of
Compton-thick absorption cannot be excluded by such variability.
Due to the limited photon statistics of the \nustar\ and/or 
previous \chandra\ and \xmm\ data, we cannot constrain 
the variability of the spectral shape (e.g., $\Gamma_{\rm eff}$)
for these three targets.

The fraction of intrinsically X-ray weak
quasars among our sample objects
is likely high. The soft stacked signal 
is not dominated by one single object,
and a lower limit on the fraction is $33\%$
(at least two out of six being intrinsically X-ray weak) with an upper limit
of 100\% (all being intrinsically \hbox{X-ray} weak).
Our targets were selected to be significantly \hbox{X-ray} weak in the
$<10$~keV band, and thus the intrinsic X-ray weakness fraction
among the general BAL-quasar population will be lower.
The fraction is estimated to be \hbox{$\approx17$--40\%}
in the LBQS BAL-quasar sample (L13), 
much larger than the $\la2\%$ fraction among
non-BAL quasars \citep{Gibson2008}. L13 
suggested that the disk wind in 
an intrinsically \hbox{X-ray} weak quasar might have
a large 
covering factor as it is likely easier to launch the wind when the nuclear
X-ray emission is weak.\footnote{Intrinsic X-ray weakness might also be associated with
powerful large-scale outflows as suggested by \citet{Teng2014}.}
Thus intrinsically \hbox{X-ray} weak quasars would be
preferentially observed as BAL quasars.

\begin{figure}
\centerline{
\includegraphics[scale=0.5]{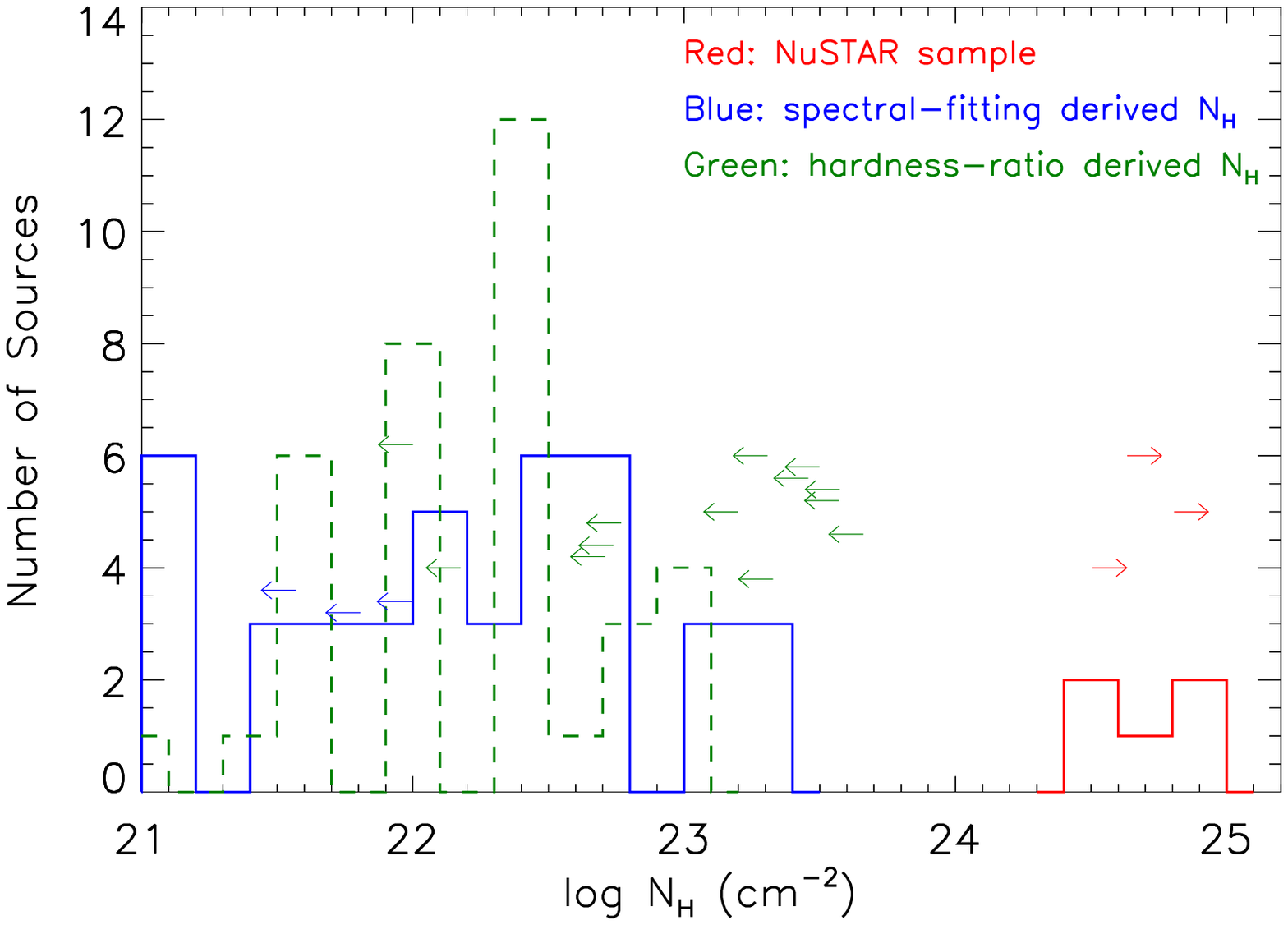}
}
\caption{
Distribution of the $N_{\rm H}$ constraints for the \nustar\
BAL-quasar sample (red histogram and lower limits), 
assuming the observed hard X-ray weakness is 
caused by absorption. For comparison, $N_{\rm H}$ distributions 
for typical BAL quasars collected from the literature
\citep{Gallagher2002a,Giustini2008,Fan2009,Streblyanska2010,Morabito2014}
are shown as the blue ($N_{\rm H}$ derived from spectral fitting) and green 
($N_{\rm H}$ derived from hardness-ratio analysis) histograms
and upper limits. Sources with $N_{\rm H}<10^{21}$~cm$^{-2}$ are
included in the $N_{\rm H}=10^{21}$~cm$^{-2}$ bin.
\label{fig-nhdist}}
\end{figure}

\section{SUMMARY AND FUTURE WORK}

We have presented \nustar\ observations of an extended sample of 
six BAL quasars with significant X-ray weakness in the $<10$~keV
band. All targets are either marginally or not detected by \nustar,
indicating significant hard X-ray \hbox{(8--24~keV)}
weakness as well, similar to the pilot sample 
in L13. The derived column-density constraints in an absorption
scenario are all in the Compton-thick regime. However, stacking 
and joint spectral analyses of the data
indicate a soft effective photon index,
generally disfavoring Compton-thick absorption. Moreover, the uniform 
hard X-ray weakness observed in this sample and also the pilot sample
suggests that the \hbox{X-ray} weakness is intrinsic in at least some of the targets.
We conclude that \nustar\ observations of BAL-quasar samples have likely
discovered a significant population ($\ga33\%$) of intrinsically 
X-ray weak sources among the 
BAL quasars with significantly weak $<10$~keV emission.
We emphasize that the disk wind in
an intrinsically \hbox{X-ray} weak quasar might have
a large covering factor, and thus the source would be
preferentially observed as an BAL quasar.

It would be worthwhile to obtain additional \nustar\ observations of
our targets, so that better constraints can be derived on the spectral 
shapes of individual and stacked sources. For example, 
by tripling the exposure times,
the three undetected sources should in general be
detected in the \hbox{3--8~keV} band and the stacked source of these three
will probably be detected in the 8--24~keV band, as suggested by
the current stacking results. Meanwhile,
PG~1001 and PG~1254 should likely be individually detected in the 
\hbox{8--24~keV} band.
The improved spectral-shape constraints would 
provide a tighter constraint on the
fraction of intrinsically
X-ray weak quasars among this sample.
Moreover, by increasing the exposure times by factors of $\approx5$--10,
sufficient photon statistics could probably be obtained 
for some individual targets to allow identification of intrinsically
X-ray weak quasars in the sample via basic spectral analysis
and accurate measurements of a soft $\Gamma_{\rm eff}$.
Such longer observations could be divided into a few 
segments to probe any short-term variability of the targets;
ideally these segments should be separated by somewhat less
than the expected light-crossing time of the shielding gas
($\approx4$--40~days in the rest frame).

It would also be valuable to select additional intrinsically
X-ray weak candidates for \nustar\ observations, although sources
satisfying the selection criteria in Section~2 
are rare largely due to the lack of 
systematic BAL-quasar selection at low redshifts ($z\la1.4$), where
the key \ion{C}{4} transition is generally not accessible via
ground-based spectroscopy.
Such studies could probably also be extended to include ``mini-BAL'' quasars,
which have narrower absorption troughs (\hbox{500--2000~km~s$^{-1}$} wide) 
than BAL quasars yet may
share their other properties \citep[e.g.,][]{Trump2006}.

Presently, the fraction of BAL quasars that are intrinsically X-ray weak
is poorly constrained in the LBQS sample (\hbox{$\approx17$--40\%}; L13),
which was derived via stacking analysis of a sample of \chandra\ 
2--8 keV undetected objects. We have scheduled additional 9--12~ks \chandra\
observations of the six undetected HiBAL quasars in the sample. 
Based on our current \chandra\ stacking results,
we suspect that these observations will convert most of these
non-detections into detections. This will set
a much tighter and more robust upper limit upon the fraction of
HiBAL quasars that are intrinsically X-ray weak. Further X-ray
observations of the undetected LoBAL quasars in the sample 
may be pursued in future work. Such studies will benefit our assessment of
the different fractions of intrinsically X-ray weak objects among
BAL and non-BAL quasars.

~\\

We acknowledge support from the California Institute of
Technology (Caltech) \nustar\ subcontract 44A-1092750 (BL, WNB),
NASA ADP grant NNX10AC99G (BL, WNB),
NASA Postdoctoral Program (SHT),
CONICYT-Chile FONDECYT 1140304 (PA) and 1141218 (FEB),
``EMBIGGEN'' Anillo ACT1101 (PA, FEB),
Basal-CATA PFB-06/2007 (FEB), 
Project IC120009 ``Millennium Institute of Astrophysics (MAS)'' of 
Iniciativa Cient\'{\i}fica Milenio del Ministerio de 
Econom\'{\i}a, Fomento y Turismo (FEB),
ASI/INAF grant I/037/12/0-011/13 (AC), 
STFC grant ST/J003697/1 (PG), and
the
Swiss National Science Foundation (NSF) grant PP00P2 138979/1 (MK).
We thank K.~Forster for help with the observation planning,
and we thank T. Yaqoob for helpful discussions.
We thank the referee for carefully
reviewing the manuscript and providing helpful comments.

This work was supported under NASA Contract No. NNG08FD60C, and made use
of data from the \nustar\ mission, a project led by Caltech,
managed by the Jet Propulsion Laboratory,
and funded by the National Aeronautics and Space Administration.
We thank the \nustar\ Operations, Software and Calibration teams for
support with the execution and analysis of these observations.
This research has made use of NuSTARDAS
jointly developed by the ASI Science
Data Center (ASDC, Italy) and Caltech (USA).


\begin{deluxetable}{lccccccccccc}

\tablecaption{\nustar\ Observation Log and Target Properties}

\tablehead{
\colhead{Object}                   &
\colhead{$z$}                   &
\colhead{$m_{B}$}                   &
\colhead{$M_{B}$}                   &
\colhead{Observation}                   &
\colhead{Observation}                   &
\colhead{Exp}                   &
\colhead{Exp\_clean}                   &
\colhead{$\Delta_{\rm OX}$}                &
\colhead{$\log M_{\rm BH}$}     &
\colhead{$\log L_{\rm bol}$}   &
\colhead{BAL}
\\
\colhead{Name}                   &
\colhead{}                   &
\colhead{}                   &
\colhead{}                   &
\colhead{Start Date}                   &
\colhead{ID}                   &
\colhead{(ks)}                   &
\colhead{(ks)}                   &
\colhead{(arcsec)}  &
\colhead{($M_{\Sun}$)}  &
\colhead{(\lum)}   &
\colhead{Type}
\\
\colhead{(1)}         &
\colhead{(2)}         &
\colhead{(3)}         &
\colhead{(4)}         &
\colhead{(5)}         &
\colhead{(6)}         &
\colhead{(7)}         &
\colhead{(8)}         &
\colhead{(9)}         &
\colhead{(10)}      &
\colhead{(11)}       &
\colhead{(12)}   
}

\startdata
PG~0043+039     &0.385&    15.9 &  $-25.5$&   2013 Jul 18&    60001119002 & 21.5&   20.2& ... &  9.0\tablenotemark{a} & 46.2 & HiBAL\\
IRAS~07598+6508 & 0.148 & 14.9 & $-24.3$& 2013 Oct 29 &60001120002 & 30.2&28.1 &...&    8.3$\tablenotemark{a}$ &46.2 &LoBAL\\
PG~0946+301 & 1.223 & 16.0 & $-28.2$ &2013 Nov 09 & 60001124002 & 37.4 & 34.9 & 6.9 & 9.8$\tablenotemark{b}$ & 47.5&HiBAL\\
PG~1001+054 &0.161&     16.1 &  $-23.2$&   2013 Jun 28&  60001122002&   20.9&   19.6&8.0 &7.7$\tablenotemark{b}$ &45.5 &HiBAL\\
PG~1254+047&1.026&      15.8& $-28.0$&  2013 Jun 27&  60001123002&   31.7&   29.4&2.9 &  9.7$\tablenotemark{b}$& 47.3 &HiBAL\\
IRAS~14026+4341 &0.323 & 15.7& $-25.3$ &2013 Nov 10 & 60001121002 & 27.8&26.0& ...&8.6$\tablenotemark{b}$ &46.3 &LoBAL
\enddata

\tablecomments{
Cols. (1) and (2): object name and redshift.
Cols. (3) and (4): apparent and absolute $B$-band magnitudes.
Cols. (5) and (6): \nustar\ observation start date and observation ID.
Cols. (7) and (8): nominal and cleaned \nustar\ exposure times, respectively.
Col. (9): minimum positional offset between the 
optical and \hbox{X-ray} positions.
The \hbox{X-ray} positions are determined from {\sc wavdetect} detections
in the 3--24 keV images of FPMA and FPMB. Blank entries indicate 
non-detections. 
Col. (10): virial BH mass from the literature; (a): \citet{Hao2005};
(b): \citet{Shen2011}. 
Col. (11): bolometric luminosity 
calculated from the scaled \citet{Richards2006}
composite quasar SED.
Col. (12): BAL quasar type depending upon whether there are BALs
from ions at low-ionization states.}
\label{tbl-obs}
\end{deluxetable}

\begin{deluxetable}{lcccccccccc}
\tablecaption{\nustar\ Photometry and Column-Density Constraints}
\tablehead{
\colhead{Object Name}                   &
\multicolumn{3}{c}{Net Counts} &
\colhead{$\Gamma_{\rm eff}$\tablenotemark{a}}                   &
\multicolumn{3}{c}{Flux ($10^{-14}$~\flux)}             &
\colhead{$\log L$}  &
\colhead{$f_{\rm weak}$\tablenotemark{b}}  &
\colhead{$N_{\rm H}$\tablenotemark{c}}  \\
\colhead{and FPM}  &
\colhead{}  &
\colhead{}  &
\colhead{}  &
\colhead{}  &
\colhead{}  &
\colhead{}  &
\colhead{}  &
\colhead{(\lum)} &
\colhead{}  &
\colhead{$(10^{24}$~cm$^{-2})$} \\
\cline{2-4}  \cline{6-8}\\
\colhead{}                   &
\colhead{3--24}                   &
\colhead{3--8}                   &
\colhead{8--24}                   &
\colhead{}                  &
\colhead{3--24}                   &
\colhead{3--8}                   &
\colhead{8--24}                   &
\colhead{3--24}    &
\colhead{3--24}    &
\colhead{} \\
\colhead{}  &
\colhead{keV}  &
\colhead{keV}  &
\colhead{keV}  &
\colhead{}  &
\colhead{keV}  &
\colhead{keV}  &
\colhead{keV}  &
\colhead{keV}  &
\colhead{keV}  &
\colhead{}
}

\startdata
PG~0043+039 A &$                    <21.8$&$                    <17.2$&$                    <12.7$&$                                      1.8$&$        <7.2$&$        <3.9$&$        <6.2$&$      <43.6$& $>7.3$& $>2.8$\\
PG~0043+039 B &$                    <18.1$&$                    <11.6$&$                    <15.3$&$                                     1.8$&$        <6.4$&$        <2.8$&$        <8.1$&$       <43.5$& $>8.4$&  $>3.2$\\
IRAS~07598+6508 A &$                    <41.4$&$                    <28.2$&$                    <21.0$&$                                    1.8$&$        <8.7$&$        <4.1$&$        <6.6$&$    <42.7$&$>22.5$& $>5.8$\\
IRAS~07598+6508 B &$                    <34.2$&$                    <27.8$&$                    <16.9$&$                                   1.8$&$        <7.6$&$        <4.2$&$        <5.6$&$     <42.6$&$>24.7$& $>6.4$\\
PG~0946+301 A &$     44.6_{-13.5}^{+15.2}$&$      21.9_{-9.2}^{+10.9}$&$      22.7_{-9.9}^{+11.6}$&$                          1.2_{-0.8}^{+0.9}$&$   8.7\pm3.4$&$   2.4\pm1.1$&$   6.2\pm3.1$&$    44.9\pm0.1$& 3.4 &$2.1_{-1.8}^{+3.6}$\\
PG~0946+301 B &$     36.4_{-14.4}^{+16.0}$&$                    <35.5$&$                    <36.8$&$                                         1.8$&$   6.4\pm2.9$&$        <4.3$&$        <9.7$&$   44.8\pm0.2$& 4.6 &$3.0_{-2.2}^{+5.0}$\\
PG~1001+054 A  &$      25.0_{-9.8}^{+11.5}$&$       17.5_{-7.2}^{+9.0}$&$                    <20.3$&$                                       >0.4$&$   8.5\pm4.0$&$   4.1\pm1.9$&$       <10.2$&$   42.8\pm0.2$& 22.7& $6.0_{-3.6}^{+10.1}$\\
PG~1001+054 B&$     26.4_{-10.8}^{+12.5}$&$       17.6_{-8.0}^{+9.7}$&$                    <20.6$&$                                       >0.4$&$   9.3\pm4.6$&$   4.2\pm2.1$&$       <10.8$&$     42.8\pm0.2$& 20.4&$5.4_{-3.3}^{+10.0}$\\
PG~1254+047 A&$     30.6_{-12.2}^{+13.9}$&$                    <27.3$&$                    <33.2$&$                                         1.8$&$   6.8\pm3.1$&$        <4.2$&$       <11.0$&$    44.6\pm0.2$&5.7& $3.4_{-2.4}^{+6.1}$\\
PG~1254+047 B&$     39.0_{-14.1}^{+15.7}$&$     27.0_{-10.4}^{+12.0}$&$                    <31.0$&$                                       >0.4$&$   9.2\pm4.0$&$   4.4\pm1.8$&$       <11.0$&$     44.7\pm0.2$&4.2& $2.5_{-1.9}^{+4.2}$\\
IRAS~14026+4341 A&$                    <20.7$&$                    <15.5$&$                    <13.7$&$                                    1.8$&$        <4.7$&$        <2.4$&$        <4.7$&$     <43.2$&$>12.5$& $>4.3$ \\
IRAS~14026+4341 B&$                    <25.9$&$                    <28.6$&$                    <11.8$&$                                     1.8$&$        <6.3$&$        <4.8$&$        <4.3$&$    <43.3$&$>9.7$& $>3.4$
\enddata
\tablenotetext{a}{Effective photon index,
derived based on the band ratio between
the observed 8--24~keV and 3--8~keV counts, assuming a power-law model with
Galactic absorption. $\Gamma_{\rm eff}=1.8$ is assumed if it cannot be
constrained.}
\tablenotetext{b}{Factor of X-ray weakness in the 3--24 keV band,
$f_{\rm weak}={F_{\rm expected}}/{F_{\rm observed}}$. The expected flux was 
derived based on the expected $\alpha_{\rm OX}$
(from the \hbox{$\alpha_{\rm OX}$--$L_{\rm 2500~{\textup{\AA}}}$} relation)
and a power-law X-ray spectrum with $\Gamma=1.8$. The scatter of the 
expected flux was accounted for when deriving the lower limit for 
an undetected source.}
\tablenotetext{c}{Absorption column-density constraint derived
from the factor of 3--24 keV weakness using the {\sc MYTorus} model.}
\label{tbl-pho}
\end{deluxetable}

\end{document}